\documentclass[preprint,3p,authoryear]{elsarticle}

\journal{Planetary \& Space Science}

\usepackage{natbib}

\usepackage{amssymb}

\usepackage{doi}
\usepackage{textcomp}
\usepackage{color}
\usepackage{ulem}
\usepackage{hyperref}
\hypersetup{colorlinks=true, urlcolor=blue}

\begin{document}

\begin{frontmatter}



\title{Evidence for Surface Variegation in Rosetta OSIRIS Images of Asteroid 2867 Steins\tnoteref{label1}\tnoteref{label2}}
\tnotetext[label1]{\doi{10.1016/j.pss.2010.04.020}}
\tnotetext[label2]{\copyright 2017. This manuscript version is made available under the CC-BY-NC-ND 4.0 licence:\\ \url{https://creativecommons.org/licenses/by-nc-nd/4.0/}}


\author[label1]{S.E. Schr\"oder}
\author[label1]{H.U. Keller}
\author[label2]{P. Gutierrez}
\author[label1]{S.F. Hviid}
\author[label1]{R. Kramm}
\author[label2]{W. Sabolo}
\author[label1]{H. Sierks}

\address[label1]{Max-Planck-Institut f\"ur Sonnensystemforschung, Katlenburg-Lindau, Germany}
\address[label2]{Instituto de Astrof\'isica de Andaluc\'ia-CSIC, Granada, Spain}

\begin{abstract}
The OSIRIS camera onboard Rosetta successfully acquired images of asteroid 2867 Steins through a variety of color filters during the flyby on 5 September 2008. The best images of this 5 km diameter asteroid have a resolution of 78 meters per pixel. We process the images by deconvolving with the point spread function and enlarging through the Mitchell-Netravali filter. The enhanced set is analyzed by means of various techniques (PCA, band ratios, stereo anaglyphs) to study surface morphology and search for variegation. We identify a landslide, which supports a YORP origin for Steins' unusual diamond shape. In addition, we find that the interior of one of two large craters on the south pole is bluer than the rest of the body.
\end{abstract}


\end{frontmatter}



\section{Introduction}
\label{sec:introduction}

On 5 September 2008 the ESA Rosetta spacecraft acquired images of E-type asteroid 2867 Steins with the on-board OSIRIS camera \citep{K09}. OSIRIS consists of a narrow angle camera (NAC) and a wide angle camera (WAC), details of which are described by \citet{K07}. The WAC performed flawlessly throughout the flyby while the NAC went into safe mode just before closest approach (Accomazzo et al., this issue), limiting the highest image resolution to 78 meters per pixel. In view of the small diameter of this asteroid (5~km) we would like to study surface features as small as 100~m, i.e.\ on the scale of a single pixel. To stretch the limits of the image resolution we enhance the images by sharpening and enlarging. We sharpen by by deconvolution with the point spread function (PSF) and enlarge through application of a filter with excellent reconstruction properties \citep{MN88}. We analyze the enhanced images by means of (1) principal component analysis, (2) band ratio images, and (3) stereo anaglyphs to study the asteroid surface variegation, morphology, and shape.

The flyby trajectory was well-suited for the search for variegation. On approach a large part of Steins's surface was illuminated while the phase angle decreased only slowly. The asteroid was imaged through a wide variety of narrow band filters from the UV into the near-IR. Previous asteroid flybys have revealed color variations on S-type asteroids Gaspra and Ida that were attributed to space weathering \citep{H94,V96}. It is thought that space weathering leads to reddening of the asteroid surface by action of the solar wind and micrometeorite impacts (e.g.\ \citealt{V09}). Not all asteroids show evidence of space weathering, for example no variegation was found for C-type asteroid Mathilde \citep{V97}, so the study of variegation on an asteroid of the rare E-type may lead to further understanding of the phenomenon.

\section{Methods and Observations}
\label{sec:observations}

\subsection{Image processing}

\begin{figure}
\centering
\includegraphics[width=8cm,angle=0]{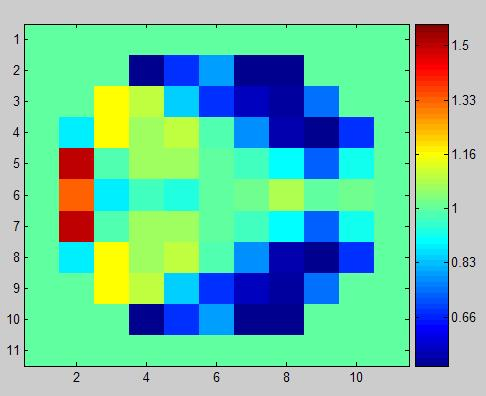}
\caption{Numerical correction map, to be multiplied with the Moffat fit for the PSF of the NAC {\it FFP-Vis/Blue} filter.}
\label{fig:psf_cor_map}
\end{figure}

Details of the OSIRIS images we selected for analysis are provided in Tables~\ref{tab:nac_images} and \ref{tab:wac_images}. The images of Steins are typically small (Steins is sized $65\times50$ pixels in the highest resolution images) and noticeably affected by the PSF. To allow a comparison on a sub-pixel scale we deconvolve the images with the PSF. Ray tracing simulations of the WAC optics have established that the PSF full width at half maximum (FWHM) is smaller than half a CCD pixel width \citep{D04}. The actual PSF under flight conditions is larger because of instrument tolerance and lateral diffusion of photo-electrons in the CCD. Before detailing how we model the PSF we briefly describe the OSIRIS filter system. Characteristics of the filters used for the images described in this paper are summarized in Table~\ref{tab:filters}; further details are provided by \citealt{K07}. In the following we write filter names in {\it italics}. A NAC image is acquired through pairs of filters in separate wheels. One of these is a color filter, whereas the other adjusts the focus (Near Focus Plate {\it NFP} or Far Focus Plate {\it FFP}) or reduces the flux ({\it Neutral}). The WAC employs a single filter. To successfully deconvolve the images we need to know the PSF of all filter combinations. We have started a still on-going study to characterize the instrument PSF and to monitor possible changes in time. We determined the PSF for several NAC and WAC filter combinations from 17 images of star fields and 28 images of Vega acquired during the calibration campaign. First, we fitted five analytical distributions (Gaussian, Lorentz, Sersic, Moffat, and Waussian) to the stellar profiles and inspected the residuals for regular patterns. In 97\% of the cases the Moffat profile achieved the best fit. The Moffat distribution is characterized by amplitude $A$ and a FWHM of $2 \sigma$ \citep{M69}:
\begin{equation}
f_{\rm M}(r) = \frac{A}{1 + (r / \sigma)^4},
\label{eq:Moffat}
\end{equation}
with $r$ the distance to the profile center (in pixels). The best fit values for $\sigma$ are listed in Table~\ref{tab:psf} for the filters investigated. The fit can be improved by adopting numerical correction maps to correct for PSF asymmetry, an example of which is shown in Fig.~\ref{fig:psf_cor_map}. We find no significant variation of the PSF across the field of view of radiometrically calibrated images.
The filter combinations used during the Steins campaign are not identical to the ones in Table~\ref{tab:psf}. While Steins was imaged through the NAC {\it Blue}, {\it Green}, and {\it Orange} filters, it was in combination with the {\it Neutral} instead of the {\it FFP-Vis} filter. This slightly alters the wings of the PSF, but not the core. We construct a Moffat PSF for these three color filters using the values in Table~\ref{tab:psf}, including the correction maps. For the other NAC filters we use the {\it Orange} PSF. The WAC filters in Table~\ref{tab:psf} were not used at Steins, so we construct a generic WAC Moffat PSF with $\sigma = 0.7$, the average of the two values in the table. We deconvolved radiometrically calibrated images (level~2), all scaled to the same brightness level, with the PSF using the method of \citet{S08}. Note that using a generic WAC PSF will lead to differences in image sharpness for the different filters.

\begin{figure}
\centering
\includegraphics[width=\textwidth,angle=0]{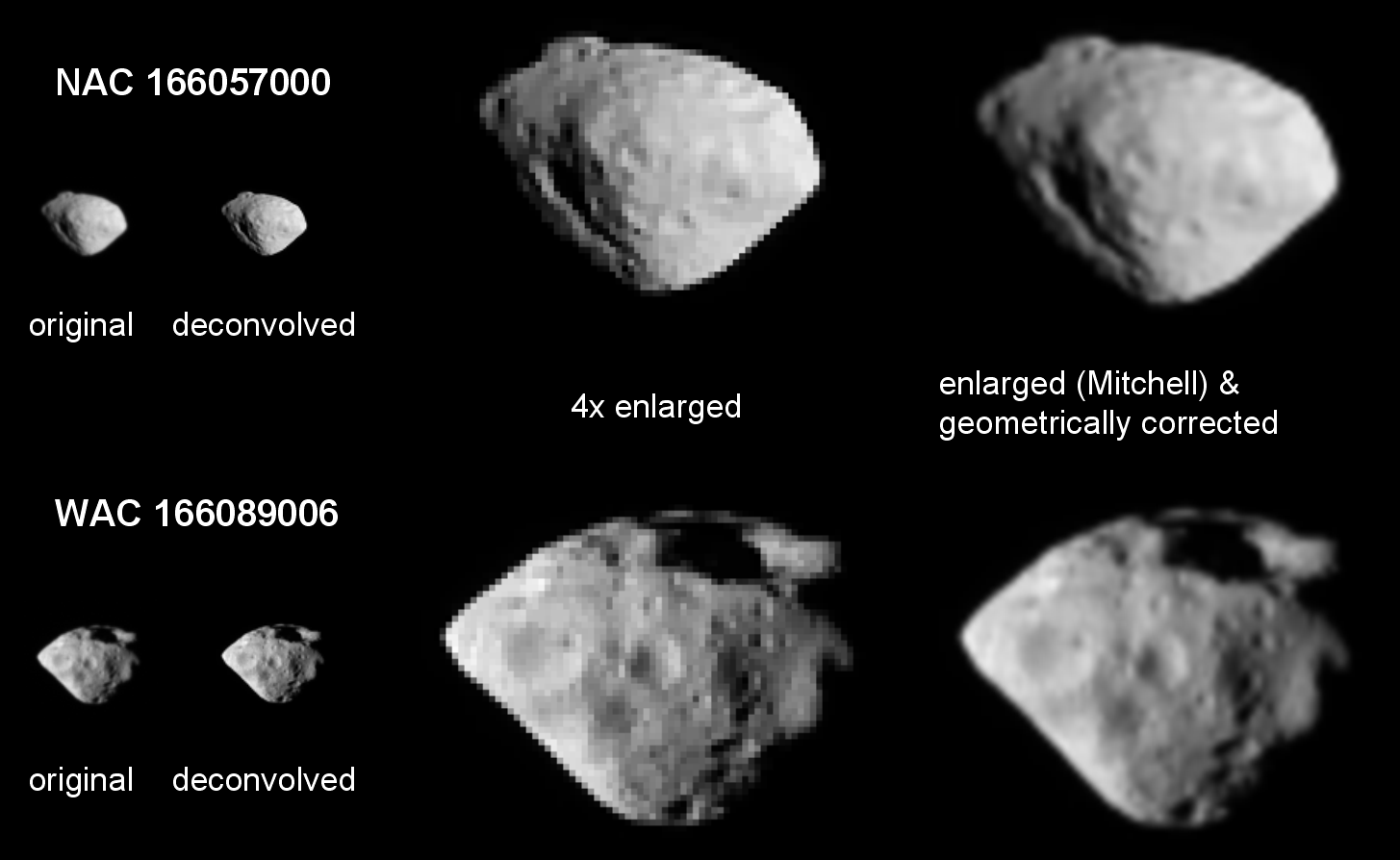}
\caption{Examples of the processing performed in this paper on both NAC and WAC images. We deconvolved and enlarged the original level~2 images to arrive at the images at far right. A fourfold enlargement of the deconvolved image using nearest-neighbor sampling shows the pixel size of the original image (center). The images shown in this paper are enlarged four times by means of the Mitchell filter, simultaneously correcting for geometric distortion.}
\label{fig:image_proc}
\end{figure}

The images were subsequently enlarged about four times by means of the cubic Mitchell-Netravali filter \citep{MN88}. This filter is optimized to reconstruct the original scene and reduce artifacts like ringing and aliasing by tuning the two parameters $B$ and $C$. We obtained good results with $B = 0.4$ and $C = 0.3$, in accordance with the $2B + C = 1$ relation recommended by \citeauthor{MN88}. While resizing we simultaneously correct for geometric distortion and varying distance to Steins. We verified that the filter does not introduce spurious artifacts by comparing the highest resolution NAC image with a filtered WAC image acquired closely in time. Figure~\ref{fig:image_proc} illustrates the full procedure. Since Steins' rotation is retrograde, its south pole is pointed towards celestial north. Following \citet{K09}, we display all images in this paper with the south pole up. The last step is registration of the images. This was done by shifting image pairs pixel-by-pixel to minimize their (absolute) difference. The accuracy thus achieved is approximately a quarter pixel of the original image.

\subsection{Methods of analysis}

The first technique we employ to study possible surface variegation on Steins is the principal component analysis (PCA). PCA can help to identify correlations in sets of images taken through different filters (bands). It effectively reduces a large number of spectral bands to a few ``principal components'' (PC) that contain most of the variance present in the full set. There are as many PCs as there are bands, and they are ordered according to their contribution to the total variance. The first PC contains that what correlates most over all bands: the overall brightness of the surface. Note that this is not necessarily equal to the reflectance or albedo, as it includes the effects of shadows and shading. Thus, PC~1 will be most similar to an actual image of the asteroid. PC~2 contains that what correlates most over all bands in a direction orthogonal to that of PC~1. It contains information about the color variation over the surface, and, in our case, will also show artifacts of the relative rotation between images. The highest order PCs are dominated by noise and artifacts, and can often be ignored.

We select sets of images recorded closely in time and through different filters for PCA. During the flyby the position of Rosetta in the sky as seen from a point on the surface of Steins changed over time. While this allows us to create stereo anaglyphs, it may prevent accurate image registration for PCA. We refer to the angular change of this position vector between images as the {\it change in viewing angle} $\Delta \varphi$. Due to the particular flyby geometry (Accomazzo et al., this issue) and the fact that the rotation rate of Steins (6.0~h) is long compared to the fly-by time (only 2 minutes passed between opposition and closest approach), the difference in viewing angle between successive images is essentially equal to the difference in phase angle. For an accurate PCA we would like all images in the set to have the same viewing angle, so we restrict ourselves to sets for which $\Delta \varphi$ between the first and the last image is small. In practice we are restricted to images acquired on approach, when Steins was fully lit and the phase angle decreased slowly. We select two NAC image sets (\#1 and \#2; Table~\ref{tab:nac_images}), the first having more bands and the second having higher spatial resolution, and one WAC image set (\#3; Table~\ref{tab:wac_images}), which has a large number of bands but a low resolution. To find $\Delta \varphi$ we calculate the phase angle difference between the first and last image in each set, and add a correction for Steins' intrinsic rotation over the relatively long time span between their acquisition. For sets~1, 2, and 3 we find $\Delta \varphi = 0.7^\circ$, $0.4^\circ$, and $2.5^\circ$, respectively, with an accuracy of $0.1^\circ$.

For WAC images taken at closer range the phase angle changed too rapidly with time to allow for a meaningful PCA, so we construct band ratio images by dividing images into pairs acquired closely in time and through different filters. Rather than getting an overview of the variegation over the full spectral range as for PCA, the ratio images tell us, for example, how the spectral slope or band depth changes over the surface. Also here, the change in viewing angle between the images of the pair leads to artifacts. We select WAC images acquired in filters {\it OI} versus {\it UV295/325} to emphasize changes in spectral slope over the 300-600~nm range.

The rapid change in viewing angle around closest approach is ideal for the construction of stereo images. We select several WAC filter {\it OI} image pairs with $\Delta \varphi$ up to $12^\circ$ to create anaglyphs with different degrees of perceived depth. In addition, we combine three of the highest resolution NAC images to generate two anaglyphs of different depth. These are cyan-red anaglyphs that can be observed through both blue-red and green-red glasses. For all stereo anaglyphs we verified (from the SPICE kernels) that the difference in viewing angle equals the difference in phase angle within $0.01^\circ$.

\section{Analysis}
\label{sec:analysis}

\subsection{Anaglyphs}

\begin{figure}
\centering
\includegraphics[width=10cm,angle=0]{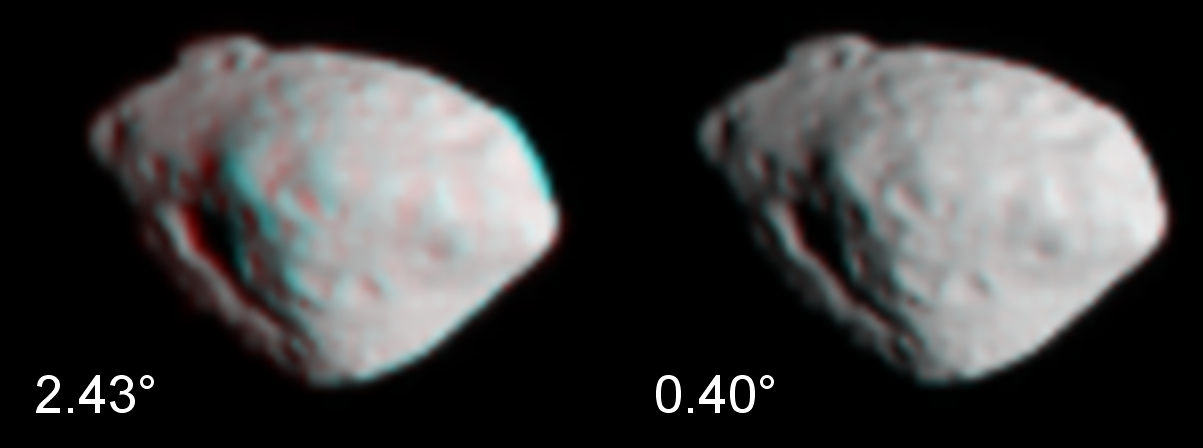}
\caption{Anaglyphs of NAC {\it Neutral/Orange} filter images. {\bf Left}: Images 166050002 and 166057004. {\bf Right}: Images 166057000 and 166057004. The angular separation between the images in the stereo pair is indicated.}
\label{fig:nac_anaglyphs}
\end{figure}

\begin{figure}
\centering
\includegraphics[width=\textwidth,angle=0]{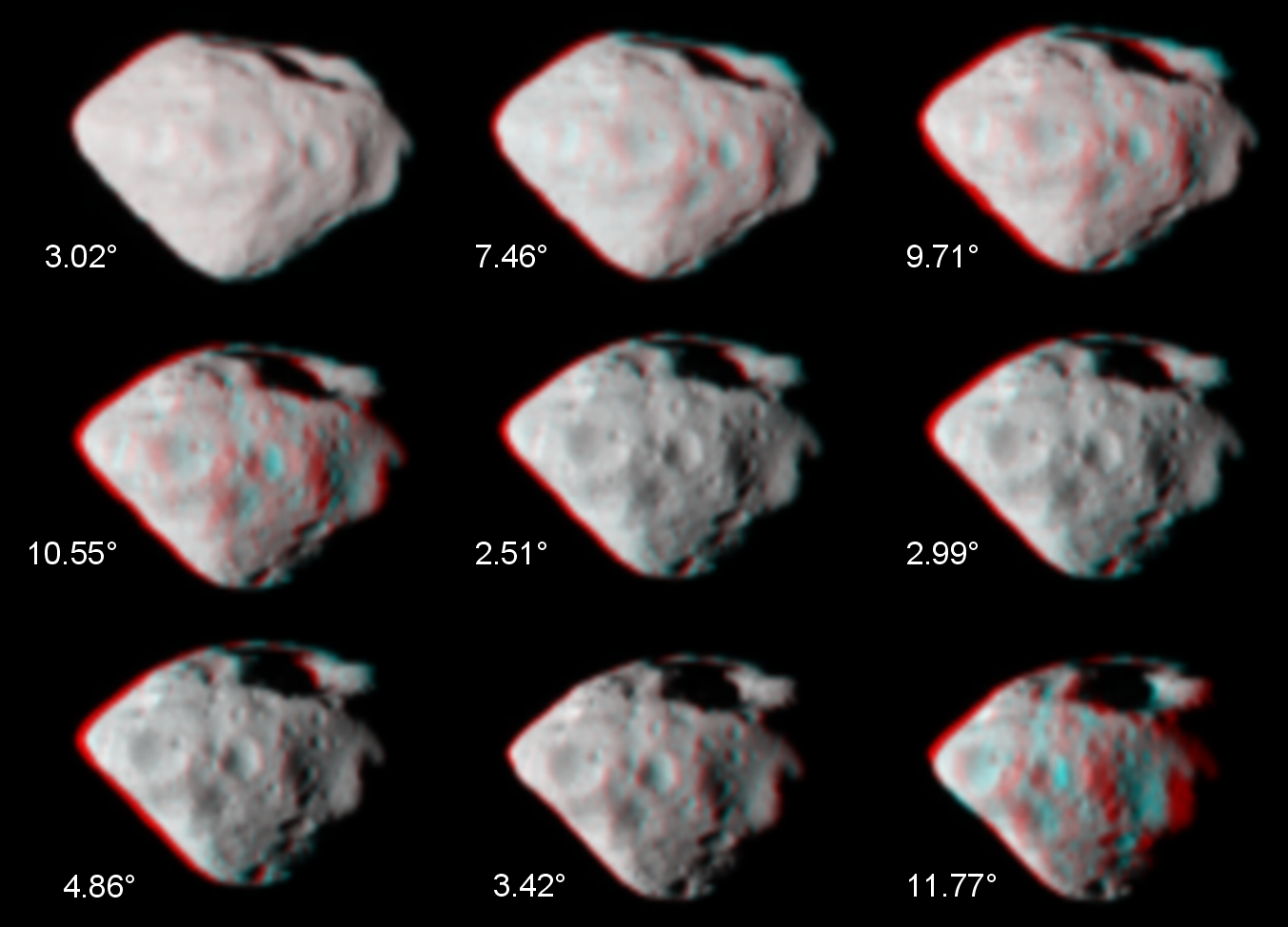}
\caption{Anaglyphs of WAC images shown at the same scale. The angular separation between the images in the stereo pair is indicated. These are anaglyphs of all consecutive {\it OI} filter image pairs between 166077025 and 166092000 (Table~\ref{tab:wac_images}).}
\label{fig:wac_anaglyphs}
\end{figure}

\begin{figure}
\centering
\includegraphics[width=\textwidth,angle=0]{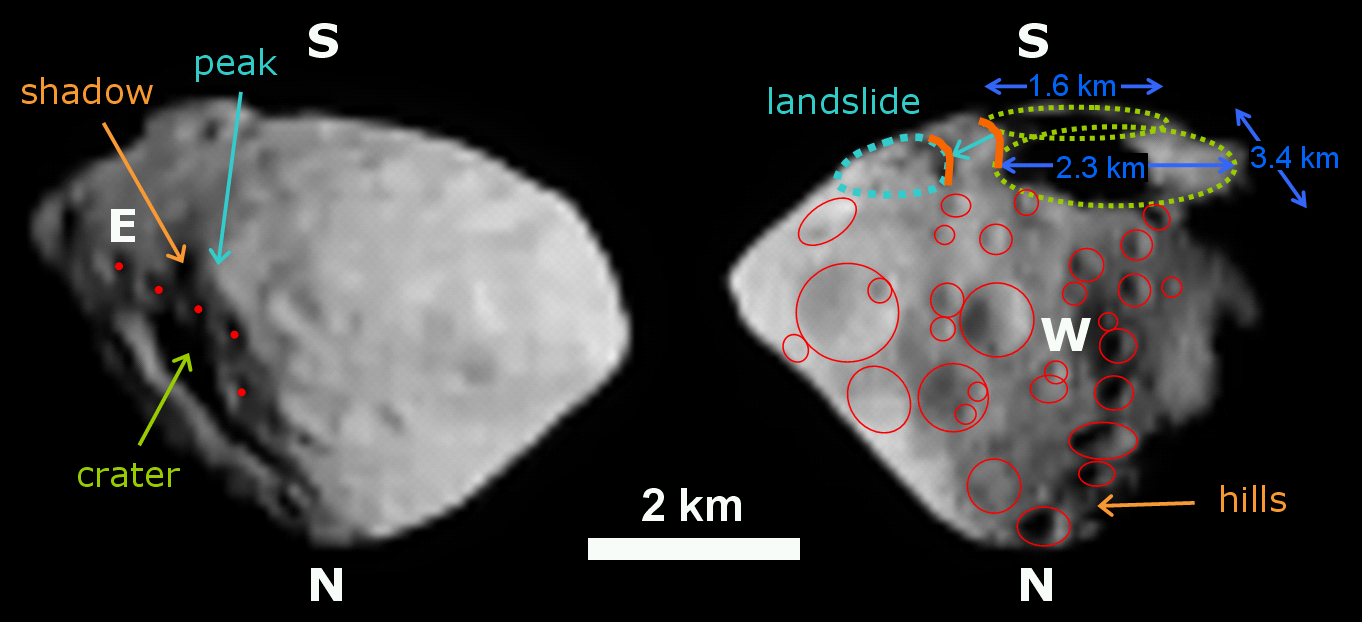}
\caption{Annotated versions of high resolution NAC and WAC images of Steins shown on the same scale. {\bf Left}: NAC image 166057004, showing the eastern side and front. The red dots indicate the location of what may be five similarly sized craters, four of which line up. A large crater (labeled) is located on the eastern side. {\bf Right}: WAC image 166089009, showing the western side. Encircled are all craters we recognize in the image. Two large craters (green dotted lines) top the south pole. The light blue dashed line outlines a landslide. The drawn orange lines emphasize the similarity in shape of the crater rim and the back of the landslide; the arrow points in the slide direction.}
\label{fig:annotated}
\end{figure}

The images reveal the shape of asteroid Steins to be that of a brilliant cut diamond with a bulging equator. We present anaglyphs constructed from NAC and WAC images in Figs.~\ref{fig:nac_anaglyphs} and \ref{fig:wac_anaglyphs}, and an annotated version of two key images in Fig.~\ref{fig:annotated}. The NAC and the (high-resolution) WAC images essentially show two different sides of Steins. We refer to the side most clearly visible in the NAC images as the {\it eastern side}, and to the side visible in the WAC images as the {\it western side}. In between these is an area where illumination is almost perpendicular, to which we refer as the {\it frontal part} of Steins. The two NAC anaglyphs in Fig.~\ref{fig:nac_anaglyphs} show us the front and the eastern side, the latter obliquely illuminated. The small angular separation between the images results in a limited depth perception. The left image shows how a mini-Steins would appear if it were located 2~m in front of our eyes; for the right image this distance is 12~m. To appreciate the resolution of the best NAC and WAC images, consider that the eye would resolve similar details on a $0.8\times1.5$~cm sized pebble at 2~m distance. Most prominent on the eastern side is a large hill on the equator that casts a shadow on the surface behind it (Fig.~\ref{fig:annotated}). Directly below this hill is a chain of what appear to be four identically sized craters with $\sim$400~m diameter. Below this putative crater chain is a large crater or chasm, of which the bottom cannot be seen. It has an irregular and elongated shape. While the illumination of this side of the asteroid is oblique, the crater wall is bright, indicating that it is steep. If created by impact, the angular corners must be the result of collapse, revealing the existence of underlying faults.

The WAC anaglyphs in Fig.~\ref{fig:wac_anaglyphs} show the western side of Steins and more of the frontal part. The anaglyphs have different perceived depths due to the different stereo angles; the distance to Steins perceived by our eyes ranges from 2~meters for the center anaglyph to 40~cm for the one at bottom right. Surface features are subdued on the front, where illumination is nearly perpendicular. A few large craters can be distinguished, but small craters, if present, are invisible. Craters, both degraded and fresh, are well visible on the western side; we count 32 craters in Fig.~\ref{fig:annotated}. Due to stronger shadows they are most pronounced near the terminator, where we find another possible crater chain. Several more equatorial hills are present. Especially well visible in the last anaglyph in Fig.~\ref{fig:wac_anaglyphs} is the double nature of the large south polar crater complex. From images 166089010 and 166089009 we measure their diameters as $2.3\pm0.1$~km (larger crater), $1.6\pm0.1$~km (smaller crater), and $3.4\pm0.2$~km for the complex (the craters partially overlap as indicated in Fig.~\ref{fig:annotated}). Located at the left edge of the crater duo in Fig.~\ref{fig:annotated} is a complex of hills, the shape of which suggests it was formed in a landslide off the crater rim. The lower part of this positive relief feature appears to be crumpled, and its upper edge follows the shape of the crater rim. The slope bridging hills and rim appears to be devoid of craters, even though the illumination favors their identification, suggesting that the landslide was a relatively recent event. Around the north pole we find a heavily eroded crater and some fresh ones, one of which is partly overlapped by hills, possibly ejecta from a nearby crater.

\subsection{PCA}

\begin{figure}
\centering
\includegraphics[width=\textwidth,angle=0]{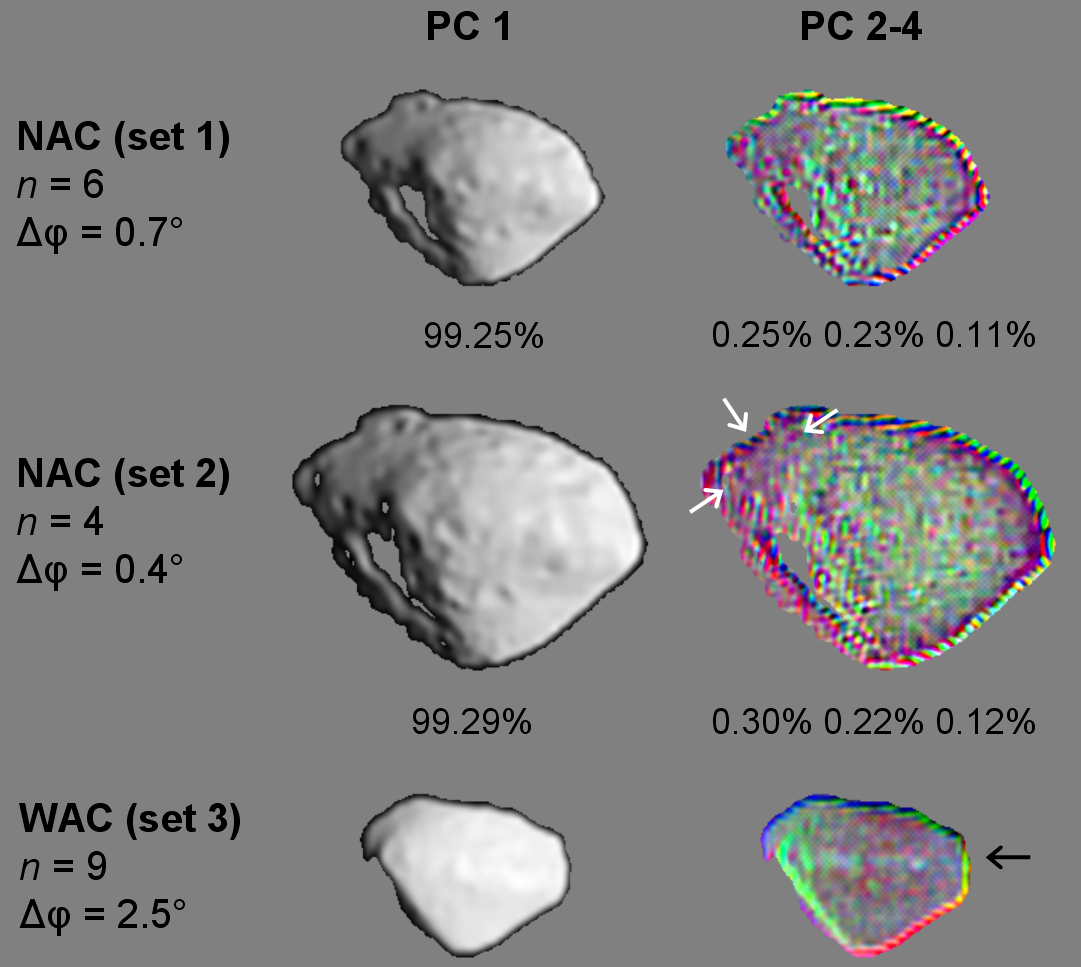}
\caption{Principal component analysis of selected NAC and WAC image sets acquired on approach (image details in Table~\ref{tab:nac_images} and \ref{tab:wac_images}). A set consists of images taken through different filters (bands) and closely in time, so that the difference in viewing angle between the first and last image in the set ($\Delta \varphi$) is limited. The left image column shows the first principal component, which contains around 99\% of the variance in the bands (exact percentage indicated). The right column shows color composites of principal components 2-4 (red: PC~2, green: PC~3, blue: PC~4; the percentage of variance explained by each PC is indicated). Image pixels not included in the PCA are neutral gray. Most of the signal in the higher principal components is due to a change in viewing angle. The flank (white arrows) appears to have a slightly different color than the front. A horizontal band across the surface in the WAC PC 2-4 image (black arrow) is a calibration artifact.}
\label{fig:pca}
\end{figure}

We performed a PCA on one WAC and two NAC sets of images acquired on approach to Steins (image details in Tables~\ref{tab:nac_images} and \ref{tab:wac_images}). The selected NAC sets offer higher spatial resolution, the WAC set a higher spectral resolution. For the NAC sets the change in viewing angle during acquisition of the image set amounts to around a quarter of a pixel; for the WAC set it is a little less than one pixel. Small as these numbers are, the effects of change in viewing angle are expected to show up in the principal components. The PCA results are shown in Fig.~\ref{fig:pca}. For all three sets PC~1 contains around 99\% of the total variance, implying an very small degree of color variegation on Steins. We display PC 2-4 in the form of an RGB image with each PC assigned to a different color channel, and each channel scaled identically to emphasize variations over the surface (allowing color saturation at the limb). While their contribution to the total variance is less than one percent, the higher order PCs are dominated by limb artifacts due to change in viewing angle. That means that any variance due to variegation must be even smaller. There are hints of variegation, though. In the NAC PC 2-4 composites the frontal side of Steins has a slightly different color than the eastern side (arrowed in Fig.~\ref{fig:pca}). This difference may result from the larger incidence and reflection angles on the side (e.g.\ \citealt{V97}), or perhaps reveal a subtle dichotomy between hemispheres. The WAC PC 2-4 are completely dominated by artifacts due to the change in viewing angle, and the only apparent color variegation is due to a calibration artifact. Imperfections in the flat field of several WAC color filters slightly darken small groups of pixels generally aligned in the CCD column direction. None of these (tiny) imperfections have been identified for the {\it OI} filter. No apparent color variation can be associated with surface features on the scale of the resolution (200-300~m).

\subsection{Opposition image}

\begin{figure}
\centering
\includegraphics[width=10cm,angle=0]{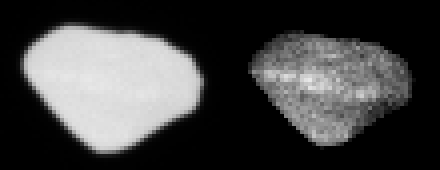}
\caption{Steins opposition image (WAC image 166077007). {\bf Left}: Image displayed with the full dynamic range. {\bf Right}: Same image, now shown with the full dynamic range assigned to pixels occupying the top 10\% of the brightness scale. These are level~2 images enlarged fourfold by nearest-neighbor sampling.}
\label{fig:opposition}
\end{figure}

The trajectory was chosen such that Rosetta would pass through solar phase angle zero (the minimum angle achieved was $0.27^\circ$). Around this crucial moment the WAC acquired a series of images in rapid succession through the {\it OI} filter. The minimum phase angle at which an image was acquired was $0.36^\circ$ (Table~\ref{tab:wac_images}), close enough to zero to fully characterize the opposition effect. In the complete absence of shadows any surface variegation should be clearly distinguishable in the opposition image. Unfortunately, we lack color information as it was not possible to acquire images through more than one filter in this fleeting moment. On first glance the close-to-zero phase angle image of Steins appears featureless (Fig.~\ref{fig:opposition}), but under closer scrutiny shows brightness variations of almost 10\%. The distribution of these variations suggests that they are not due to variegation. The brightest patches can be found on the equator in the disk center, the darkest ones close the limb. Considering that Steins is roughly diamond shaped, the incidence and reflection angles ($\iota$ and $\epsilon$) are small on the equator and large towards the limb. Thus the opposition image is essentially limb-darkened. The surface in the bright patches on the equator is perpendicular to the line of sight, and is the consequence of the undulating series of hills or craters present there. Limb darkening, a multiple scattering phenomenon, can be described by the Minnaert photometric model:
\begin{equation}
\mu I = I_0 (\mu_0 \mu)^{k},
\label{eq:minnaert}
\end{equation}
with $\mu_0 = \cos \iota$ and $\mu = \cos \epsilon$. The intensity $I$, and parameters $I_0$ and $k$ are functions of phase angle $\alpha$ and wavelength $\lambda$. At opposition $\mu = \mu_0$, and Eq.~\ref{eq:minnaert} reduces to $I = I_0 \mu^{2k-1}$. A Lambert surface has $k = 1$, whereas $k = 0.5$ equals an absence of limb darkening. We estimate the degree of limb darkening for Steins from the opposition image in Fig.~\ref{fig:opposition} by fitting Eq.~\ref{eq:minnaert} to intensity profiles across the equator, and find $k(\alpha = 0^\circ) = 0.54\pm0.01$ at 632~nm (the equator in the \citealt{K09} shape model is roughly circular). Steins' limb darkening at opposition is stronger than that for the Moon, which has virtually no limb darkening at 445~nm \citep{HV87}. For Mars, \citet{GP95} find $k(0^\circ)$ to depend on wavelength and geometric albedo ($a$). Steins' value is similar to that found for low-albedo Martian terrain at visible wavelengths ($k \sim 0.55$ for $a<0.2$), but much less than that for terrain of high albedo ($k \sim 0.75$ for $a=0.4$). \citet{K09} report a geometric albedo of $a = 0.40\pm0.01$ for Steins at 632~nm.

\subsection{Band ratio images}
\label{sec:band_ratios}

\begin{figure}
\centering
\includegraphics[width=\textwidth,angle=0]{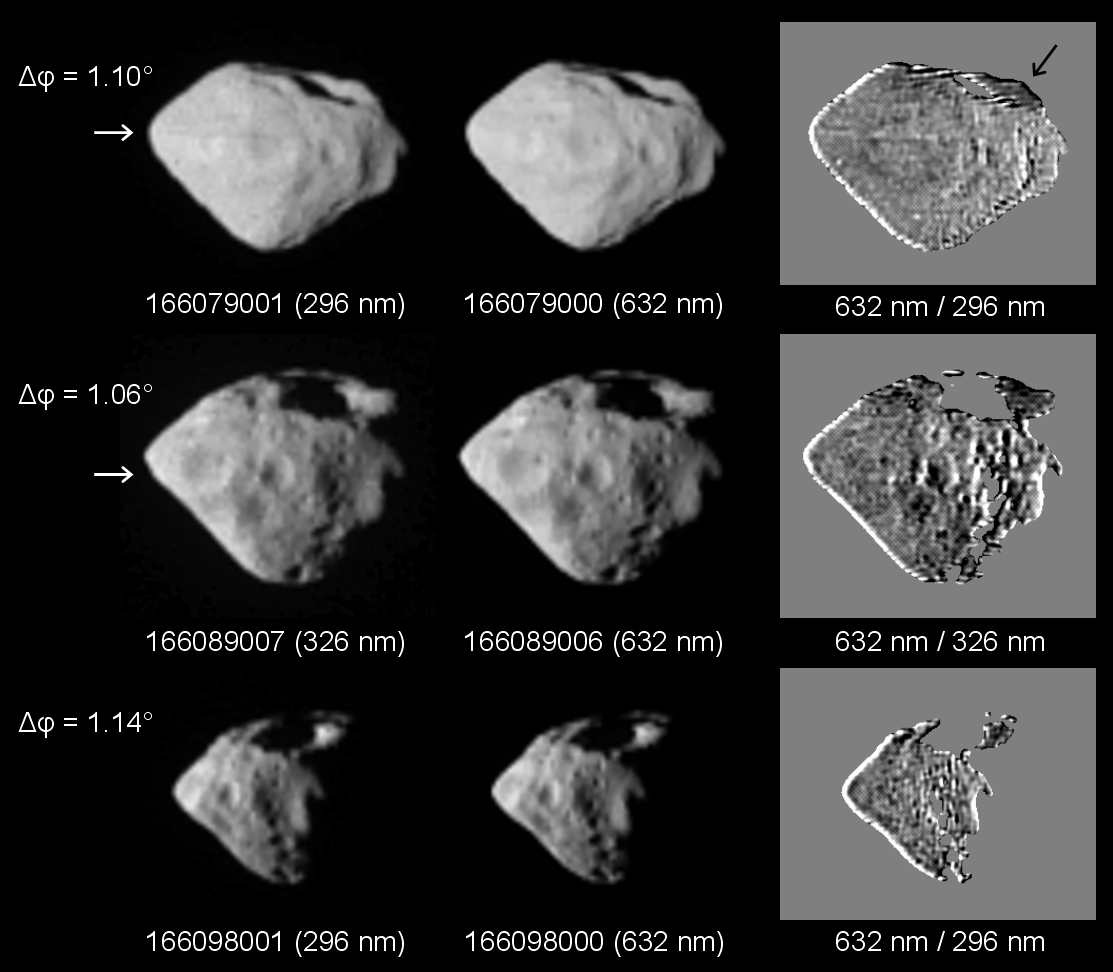}
\caption{WAC ratio images. Shown are three image pairs that were divided to obtain the (normalized) ratio images in the third column, in which black and white are assigned to ratios that are 15\% lower and higher than average, respectively. Indicated are the image numbers and the central wavelength of the filter. Due to the small change in viewing angle between the images ($\Delta \varphi$), the image pairs form stereo pairs. Relatively dark horizontal lines (white arrows) that show up bright in the ratio images are associated with a calibration artifact. The interior of the large crater (black arrow) has a 5-10\% lower ratio than average in all three image pairs. The phase angle difference between the top and bottom image pairs is $67^\circ$.}
\label{fig:ratio}
\end{figure}

As the NAC had entered into safe mode only the WAC obtained images around closest approach (Table~\ref{tab:wac_images}). At this stage the change in viewing angle was too rapid to allow for an accurate registration of images; consequently, these mismatches would completely dominate a PCA. Therefore, we concentrate on band ratio images. The WAC filters do not cover the 0.5~$\mu$m absorption band, so we restrict ourselves to investigating variations in spectral slope by comparing images in the red and ultraviolet part of the spectrum. For this we select filters {\it OI} (central wavelength 632~nm), {\it UV295} (296~nm), and {\it UV325} (326~nm). Figure~\ref{fig:ratio} shows three image pairs acquired before, around, and after closest approach. The frontal part of Steins appears bland in the ratio images. Regions of contrast are mostly confined to the terminator region, where shadows are prominent. Since these high-contrast features have mostly a vertical orientation they must be due to the difference in viewing angle between the images and not variegation, even though this difference is only around $1^\circ$. Small differences in the PSF between the filters also lead to artifacts in the ratio (the deconvolved {\it OI} images are sharper than the {\it UV295/325} images). In short, any structure in the ratio images on the same scale as the artifacts on the limb may not be real. However, one area stands out as having a distinctly different color from the bulk of the body; the inside of the largest crater on the south pole is 5-10\% more blue in all three ratio images. This area is large in terms of number of pixels, so artifacts due to differences in viewing angle between the images are not expected to dominate.

\section{Discussion}
\label{sec:discussion}

\begin{figure}
\centering
\includegraphics[width=8cm,angle=0]{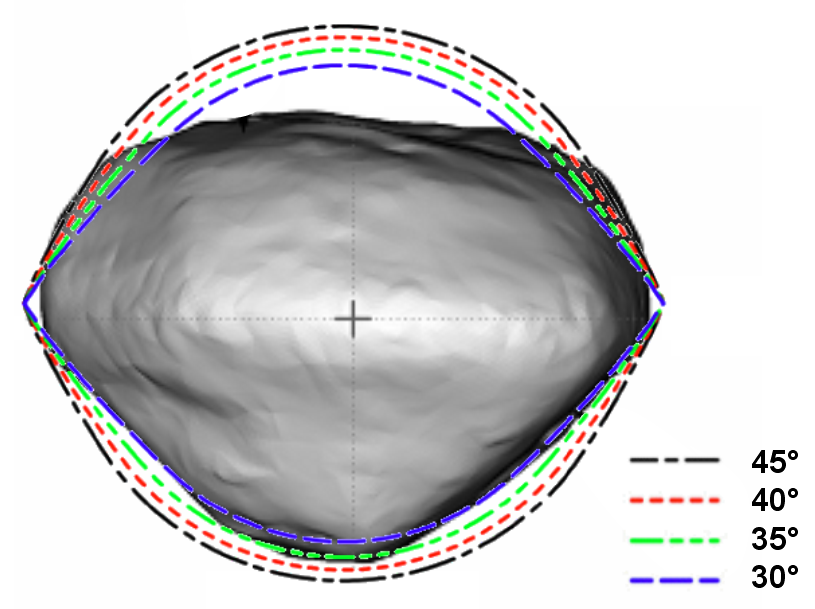}
\caption{Slope profiles for a critically spinning body \citep{H09} superposed on a shape model of Steins \citep{K09}. The numbers refer to the local slope angle.}
\label{fig:slopes}
\end{figure}

The scientific analysis of the OSIRIS images of Steins is challenging because of the limited spatial resolution. Image enhancement as described in this paper increases the amount of visible surface detail; for maximum clarity the images should be viewed in stereo. For example, Fig.~\ref{fig:wac_anaglyphs} enables us to identify a landslide not recognized previously. This is significant because \citet{K09} propose that the YORP effect is responsible for Steins' peculiar diamond shape, while \citet{H09} suggest landslides as a mechanism through which YORP reshapes asteroids. A most convincing case for YORP reshaping is provided by the 1.5~km sized near-Earth asteroid 1999 KW4 \citep{O06}. \citeauthor{H09} calculate profiles of equal local slope angle for a critically spinning body, for which the gravity equals the centrifugal force at the equator. They found a `critical slope' of $37^\circ\pm5^\circ$ for KW4. By comparing their slope profiles with the \citet{K09} shape model in Fig.~\ref{fig:slopes}, we find that the critical slope for Steins is between $30^\circ$ and $45^\circ$. The good agreement with the value for KW4 provides strong support for YORP having changed Steins' shape through landslides. \citeauthor{H09} tinkered with the polar shape to prevent the moment of inertia being greater about an axis through the equator than the axis through the poles. Just like for KW4, Steins' north pole is more pointed than predicted, which suggests that their model can be refined.

Steins appears to be a battered body, its surface punctured by large craters. \citet{K09} pointed out the presence of a large, 2.1~km diameter crater on the south pole, visible in the WAC images. When viewing the images in stereo we recognize it to be a double crater. We measure a slightly larger diameter for the larger of the two craters (2.3~km). Even though the \citeauthor{K09} shape model does not include craters (being a convex model), there is room for a double crater with the dimensions we find. Another large crater is visible in the NAC images, whose shape suggests the presence of faults in the interior. A seemingly linear chain of craters near the terminator on the WAC images was identified by \citet{K09} and Marchi et al.\ (this issue) as collapsed pits in a fault. The authors noted a relative lack of small craters after removal of these pits from the crater cumulative distribution. We identify about the same number of crater-like features as Marchi et al.\ in Fig.~\ref{fig:annotated} (right), but, unfortunately, even our enhanced images do not allow us to distinguish between pits or craters. Our crater count is most likely incomplete. \citet{W05} found the average size of craters counted on the lunar surface to be a function of incidence angle. Where sunlight strikes the surface perpendicularly (i.e.\ on the limb in Fig.~\ref{fig:annotated}), relatively small craters are not recognized, making them appear concentrated along the terminator.

Our task of identifying surface variegation is hindered by the relatively low image resolution. We could not detect color differences at the scale of the smallest craters in the highest resolution images ($\sim250$~m) due to the rapidly changing viewing angle. A PCA of the lower resolution approach images does not reveal variegation at this scale. The high degree of uniformity of Steins's surface is confirmed by the analysis of Leyrat et al.\ (this issue). We do identify subtle large scale color differences, which may be associated with different illumination conditions due to the asteroid shape. The only clear evidence for intrinsic variegation can be found on the inside of a large crater on the south pole. Here orange/UV ratio images show that the surface is 5-10\% more blue (or less red) than average. This had escaped detection by \citet{K09}, who only reported the PCA of the approach images (in which the crater is not visible). Previous observations in the visible wavelength range found the spectrum of Steins to be constant \citep{D09}. This does not necessarily conflict with our findings, as that search covered only 30\% of the rotational light curve and the interior of the large south polar craters are not well visible from Earth. \citet{W08} found one hemisphere of Steins to be significantly redder than the other over the 600-900~nm range. While we do not find evidence for such a clear dichotomy (the illuminated interior of the large crater contributes relatively little), we note that little more than half of the surface was observed by Rosetta. As mentioned above, NAC images do reveal a subtle, large scale color gradient that is probably due to different illumination conditions, not surface variegation. The WAC does not, but its filters do not cover the wavelength range for which the dichotomy was reported.

Why is the inside of the large crater bluer than the rest of the asteroid surface? As it is observed over a large range of phase angles, the bluing is not due to a different degree of phase reddening. It is also not associated with a particular combination of incidence and reflection angles because the same illuminations conditions are found on the front of the body. It must therefore be intrinsic to the surface. Perhaps it can be understood in terms of space weathering. For S-type asteroids, space weathering is thought to lead to reddening through the formation of nanophase metallic iron particles and atomic displacements in the crystal structure of silicates (see \citealt{L06} for an overview). For E-type asteroids, thought to consist of iron-poor material \citep{FL01,C04}, the consequences of space weathering are unclear. According to \citet{A08} the largest undegraded crater on an asteroid is most likely the one that has `reset' the surface, i.e.\ erased all earlier craters. On Steins, the large 2.3~km crater on the south pole would represent this `critical crater'; it should therefore be the oldest and most weathered crater on the surface, unless the large crater walls have fully protected the interior from the influx of energetic particles and micrometeorites. Of course, it is possible that the largest crater on Steins is in fact very young and of `sub-critical' size. A larger critical crater diameter would imply that Steins' interior is even more porous than that of C-type asteroid Mathilde (following Fig.~2 in \citealt{A08} and assuming gravity scaling). Alternatively, the large impact may have exposed an interior of different composition or material properties. If the interior of Steins has a different color, the ejecta of the large impact should have colored the rim. Unfortunately, even though there are hints of bluing on the rim in Fig.~\ref{fig:ratio}, the image resolution is not sufficient to make a positive identification. For now, the question of what is responsible for the different color of the large crater interior remains unanswered.

\section*{Acknowledgements}

The authors thank Qi Shan for providing a custom 16-bit deconvolution executable and an anonymous referee for providing valuable comments that have led to improvements in the manuscript.



\bibliography{Steins}

\newpage
\clearpage

\begin{table}
\centering
\caption{The Moffat $\sigma$ parameter (in pixels) from Eq.~\ref{eq:Moffat} derived for the PSF of several filter combinations.}
\vspace{5mm}
\begin{tabular}{|lll|}
\hline
\hline
camera & filter & Moffat $\sigma$ \\
\hline
NAC & FFP-Vis/FFP-IR & $0.95 \pm 0.09$ \\
    & FFP-Vis/Orange & $0.87 \pm 0.06$ \\
    & FFP-Vis/Green  & $1.05 \pm 0.04$ \\
    & FFP-Vis/Blue   & $1.03 \pm 0.06$ \\
WAC & R              & $0.68 \pm 0.05$ \\
    & Green          & $0.72 \pm 0.04$ \\
\hline
\hline
\end{tabular}
\label{tab:psf}
\end{table}

\begin{table}
\centering
\caption{Characteristics of selected OSIRIS filters used during the Steins campaign \citep{K07}.}
\vspace{5mm}
\begin{tabular}{|llcc|}
\hline
\hline
       &        & wavelength & bandwidth \\
camera & filter & (nm) & (nm) \\
\hline
NAC & Blue        & 481 & 75 \\
    & Green       & 536 & 62 \\
    & Orange      & 649 & 85 \\
    & Hydra       & 701 & 22 \\
    & Red         & 744 & 64 \\
    & Fe$_2$O$_3$ & 932 & 35 \\
    & IR          & 989 & 38 \\
WAC & UV295       & 296 & 11 \\
    & OH-WAC      & 310 &  4 \\
    & UV325       & 326 & 11 \\
    & NH          & 336 &  4 \\
    & UV375       & 376 & 10 \\
    & CN          & 388 &  5 \\
    & NH$_2$      & 572 & 12 \\
    & Na          & 591 &  5 \\
    & OI          & 632 &  4 \\
\hline
\hline
\end{tabular}
\label{tab:filters}
\end{table}

\begin{table}
\centering
\caption{Details of selected NAC images of Steins. The third column lists the time elapsed since UTC 18:00:00 on 5 September 2008 (J2000). The last column refers to the PCA set numbers in Fig.~\ref{fig:pca}. Resolution is in meters per pixel in the horizontal direction. Note that image 166057004 is the highest resolution NAC image available.}
\vspace{5mm}
\begin{tabular}{|clccccc|}
\hline
\hline
   &        & time & phase angle & distance & resolution & \\
\# & filter & (s)  & ($^\circ$)  & (km)     & (m/px)     & set \\
\hline
166050000 & Neutral/Blue       & 1459.57 & 32.20 & 7286.2 & 136  & 1 \\
166050001 & Neutral/Green      & 1463.13 & 32.17 & 7255.7 & 136  & 1 \\
166050002 & Neutral/Orange     & 1467.38 & 32.14 & 7219.3 & 135  & 1 \\
166050003 & Neutral/Hydra      & 1472.17 & 32.10 & 7178.3 & 134  & 1 \\
166050004 & Neutral/Red        & 1476.20 & 32.07 & 7143.8 & 134  & 1 \\
166050006 & Fe$_2$O$_3$/FFP-IR & 1488.42 & 31.98 & 7039.2 & 132  & 1 \\
166057000 & Neutral/Orange     & 1670.16 & 30.11 & 5486.4 & 103  & 2 \\
166057001 & Neutral/Blue       & 1674.06 & 30.06 & 5453.2 & 102  & 2 \\
166057002 & Neutral/Green      & 1677.96 & 30.01 & 5420.0 & 101  & 2 \\
166057003 & IR/FFP-IR          & 1681.85 & 29.95 & 5386.8 & 101  & 2 \\
166057004 & Neutral/Orange     & 1699.66 & 29.71 & 5235.1 & 97.9 &   \\
\hline
\hline
\end{tabular}
\label{tab:nac_images}
\end{table}

\begin{table}
\centering
\caption{Details of selected WAC images of Steins. The third column lists the time elapsed since UTC 18:00:00 on 5 September 2008 (J2000). The last column refers to the PCA set numbers in Fig.~\ref{fig:pca}. Resolution is in meters per pixel in the horizontal direction.}
\vspace{5mm}
\begin{tabular}{|clccccc|}
\hline
\hline
   &        & time & phase angle & distance & resolution & \\
\# & filter & (s)  & ($^\circ$)  & (km)     & (m/px)     & set \\
\hline
166075000 & OI     & 2125.73 & 10.41 & 1703.1 & 165  & 3 \\
166075001 & UV295  & 2128.16 & 10.08 & 1684.7 & 163  & 3 \\
166075002 & OH-WAC & 2130.65 &  9.73 & 1665.8 & 161  & 3 \\
166075003 & UV325  & 2132.97 &  9.39 & 1648.4 & 159  & 3 \\
166075004 & NH     & 2134.86 &  9.11 & 1634.2 & 158  & 3 \\
166075005 & UV375  & 2136.88 &  8.81 & 1619.0 & 157  & 3 \\
166075006 & CN     & 2138.31 &  8.59 & 1608.3 & 156  & 3 \\
166075007 & NH$_2$ & 2139.71 &  8.38 & 1597.8 & 155  & 3 \\
166075008 & Na     & 2141.06 &  8.17 & 1587.8 & 154  & 3 \\
166077007 & OI     & 2182.03 &  0.36 & 1295.6 & 125  &   \\
166077025 & OI     & 2236.13 & 17.01 & 973.45 & 94.2 &   \\
166079000 & OI     & 2243.10 & 20.03 & 940.77 & 91.0 &   \\
166079001 & UV295  & 2245.54 & 21.13 & 930.00 & 90.0 &   \\
166082000 & OI     & 2258.62 & 27.49 & 878.44 & 85.0 &   \\
166084000 & OI     & 2276.37 & 37.20 & 828.15 & 80.1 &   \\
166088000 & OI     & 2294.00 & 47.75 & 804.28 & 77.8 &   \\
166089000 & OI     & 2298.09 & 50.26 & 802.76 & 77.7 &   \\
166089003 & OI     & 2302.93 & 53.23 & 802.95 & 77.7 &   \\
166089006 & OI     & 2310.91 & 58.11 & 807.99 & 78.2 &   \\
166089007 & UV325  & 2312.66 & 59.17 & 809.87 & 78.3 &   \\
166089009 & OI     & 2316.59 & 61.53 & 815.10 & 78.8 &   \\
166092000 & OI     & 2337.37 & 73.30 & 864.54 & 83.6 &   \\
166098000 & OI     & 2368.45 & 87.76 & 995.61 & 96.3 &   \\
166098001 & UV295  & 2371.35 & 88.90 & 1010.6 & 97.8 &   \\
\hline
\hline
\end{tabular}
\label{tab:wac_images}
\end{table}

\newpage
\clearpage

\end{document}